\documentclass[manuscript]{aastex}
\usepackage{tikz}
\usepackage{amsmath}

\usepackage{graphicx}

\usepackage{pdflscape}

\slugcomment{\textbf{The Astronomical Journal} }

\shorttitle{LPCs}
\shortauthors{Jewitt}

\begin{document}

\title{Destruction of Long-Period Comets}


\author{David Jewitt$^{1}$\\
} 
\affil{$^1$Department of Earth, Planetary and Space Sciences,
UCLA, \\ 595 Charles Young Drive East, Los Angeles, CA 90095-1567, USA}
%\affil{$^2$Australia}

%\affil{$^1$Department of Earth, Planetary and Space Sciences,
%UCLA, 
%595 Charles Young Drive East, 
%Los Angeles, CA 90095-1567\\
%$^2$Department of Physics and Astronomy,
%University of California at Los Angeles, \\
%430 Portola Plaza, Box 951547,
%Los Angeles, CA 90095-1547\\
%$^3$Institut for Geophysik und Extraterrestrische Physik, Technische Universitat Braunschweig,
%Mendelssohnstr. 3, 38106 Braunschweig, Germany
%} 

\email{jewitt@ucla.edu}

\begin{abstract}
We identify a sample of 27 long-period comets for which both non-gravitational accelerations and Lyman-$\alpha$ based gas production rates are available.   Seven of the 27 comets (i.e.~$\sim$25\%) did not survive perihelion because of nucleus fragmentation or complete disintegration. Empirically, the latter nuclei  have the smallest gas production rates and the largest non-gravitational accelerations, which are both indicators of small size.  Specifically, the disintegrating nuclei have a median radius of only 0.41 km, one quarter of the  1.60 km median radius of those surviving perihelion.  The disintegrating comets also have a smaller median perihelion distance (0.48 au) than do the survivors (0.99 au).
We  compare the order of magnitude timescale for outgassing torques to change the nucleus spin, $\tau_s$, with the time spent by each comet in strong sublimation, $\Delta t$, finding that the disrupted comets are those with $\tau_s < \Delta t$.
The destruction of near-Sun long-period comets is thus naturally explained as a consequence of rotational break-up. We discuss this process as a contributor to Oort's long mysterious ``fading parameter''. 
 
\end{abstract}

\keywords{comets: general---comets: }

\section{INTRODUCTION}
\label{intro}

Comets are volatile-rich products of accretion  in the Sun's protoplanetary disk.  Historically, comets were classified as either short-period or long-period, depending on whether their orbital periods were less or greater than 200 years. With some complications and subtleties, this division into two dynamical groups has survived to the present day.  Most short-period (strictly ``Jupiter family'') comets arrive from the Kuiper belt, where they have been stored for the last 4.5 Gyr at temperatures $\sim$40 K.  Long-period comets arrive from equally long-term residence in the Oort cloud, where the equilibrium temperature is $\lesssim$10 K.  Both short and long-period comets are thought to have formed in the giant planet region of the solar system and were scattered to their respective storage reservoirs in the final stages of the growth of the planets.  

\cite{Oort50} first examined the distribution of orbital binding energies of long-period comets and found that no purely dynamical model could fit the large observed ratio of first-appearance (``dynamically new'') comets to returning comets. He invoked an ad-hoc ``fading parameter'' by which to decrease the brightness of returning comets and thereby to depress their number in magnitude limited surveys.  Subsequent work with a much larger comet sample verified both the dynamics and the need for a fading parameter \citep{Wiegert99, Levison02} but did not identify its physical origin. Observationally, \cite{Bortle91} reported that intrinsically faint long-period comets with small perihelia are less likely to survive perihelion than their brighter counterparts but, again, did not identify the mechanism.

Several practical difficulties are inherent in the observational study of long-period comets.  Crucially, these objects are generally discovered only a short time before perihelion and they are soon thereafter lost as they recede from the Sun and fade to permanent invisibility.  The observational window for long-period comets is consequently short, unlike that of the short-period comets, which can be predictably observed over many orbits.  Even worse, long-period comets with perihelia $<$ 1 au must be observed at small solar elongation (Sun-Earth-comet) angles, where ground-based telescopes struggle against low elevation and high sky-brightness constraints.  As a result of these practical difficulties, the nuclei of long-period comets are poorly characterized relative to those of short-period comets.  Cometary disintegrations, being intrinsically unpredictable and rapidly evolving, are even more difficult to study and rarely reach publication in the refereed literature.

\section{THE COMET SAMPLE}

We are interested in long-period comets for which there exist both reliable measurements of the total outgassing rates and of the non-gravitational accelerations.  For this reason, we focus on long-period comets discovered since the year 2000, to coincide with the period in which high quality physical and dynamical measurements have become routinely available from long-term surveys.  Specifically, we rely on systematic measurements of the water production rate obtained from  Lyman-$\alpha$  observations with the space-borne SWAN ultraviolet spectrometer aboard SOHO \citep{Bertaux97}, and on all-sky optical surveys (e.g.~the Catalina and, recently, Pan STARRS surveys) which provide precise astrometry over a long timebase.   We further restricted the sample to comets with perihelion distances $q \le$ 2 au.

\subsection{Non-gravitational Accelerations:} We used  orbit solutions from JPL's Horizons\footnote{\url{http://ssd.jpl.nasa.gov/horizons.cgi}} web site to compile a list of LPCs showing evidence for non-gravitational acceleration.  The orbital properties of the 27 comets used for this study are listed in  Table \ref{list}, where we show the estimated original barycentric orbital elements  in order to avoid the effects of planetary perturbations. (To find the latter, we used Horizons to compute the osculating orbital elements on 1900-Jan-01, a time at which all of the comets in the present study were far beyond the planetary region and therefore subject to minimal planetary perturbation). Twelve of the 27 comets are retrograde (inclination $i \ge$ 90\degr), consistent with being drawn from an isotropic distribution, and the eccentricities are all very close to $e$ = 1.  By default, non-gravitational parameters are computed when purely gravitational orbits fail to reproduce astrometric data, but otherwise are assumed to be zero.  Conventionally,  the non-gravitational acceleration is resolved into three, orthogonal components ($A_1$, $A_2$ and $A_3$, expressed in au day$^{-2}$), with $A_1$ being in the radial direction, $A_3$ perpendicular to the plane of the orbit and $A_2$ perpendicular to $A_1$ and $A_3$ (Marsden et al.~1973).  The radial component, $A_1$, is normally dominant, because cometary mass loss is concentrated on the heated, sun-facing side of the nucleus producing a recoil force that acts away from the Sun.  Gas produced by the sublimation of cometary water ice dominates the instantaneous outflow momentum from the nucleus.  For this reason, it is conventional to scale the acceleration by a function representing the instantaneous equilibrium sublimation rate, expressed as $g(r_H)$, such that the total acceleration is

\begin{equation}
\alpha_{NG} = g(r_H) \left(A_1^2 + A_2^2 + A_3^2\right)^{1/2}.
\label{alpha}
\end{equation}

\noindent Function $g(r_H)$ is defined by

\begin{equation}
g(r_H) = \alpha_M \left(\frac{r_H}{r_0}\right)^{-m} \left[1 + \left(\frac{r_H}{r_0}\right)^n\right]^{-k}
\end{equation}

\noindent where $r_0$ = 2.808 au, $m$ = 2.15, $n$ = 5.093, $k$ = 4.6142, and $\alpha_M$ = 0.1113 are constants determined from a fit to a model of sublimation and the normalization is such that $g(1)$ = 1 \citep{Marsden73}.  These constants derive from a model of a sublimating isothermal sphere. While this is  logically incorrect (since an isothermal sphere would sublimate isotropically and hence experience no net recoil force), the difference from a physically more plausible hemispheric sublimator is minor, at least at distances $r_H \lesssim$ 3 au where the bulk of the absorbed energy is used to break hydrogen bonds in water and the sublimation rate is large.   The assumed nucleus temperature distribution over the nucleus surface matters more at  larger distances, but the sublimation rate falls exponentially and the resulting recoil force due to distant activity is comparatively small.  Values of $A_1$, $A_2$ and $A_3$ and $\alpha_{NG}$ are listed for each comet in Table \ref{properties}.
The accelerations are small (the median value at 1 au is $\alpha_{NG} = 4\times10^{-7}$ m s$^{-2}$, about 0.007\% of the solar gravitational acceleration).

\subsection{Production Rates:} We searched the literature to find reliable measurements of the mass production rates from those long-period comets having non-zero non-gravitational accelerations.  Water molecules dominate the mass flux and outflow momentum from the nucleus. Direct measurements of the water production rate are impractical, but observations of water photodissociation products provide an accurate alternative.  We placed the greatest reliance on a long and spectacular series of measurements of Lyman-$\alpha$ emission from cometary hydrogen, a photodissociation product of water, made using the SWAN mapping spectrometer instrument on the SOHO spacecraft (Bertaux et al.~1997).   These measurements (Combi et al.~2008, 2009, 2011, 2014, 2018, 2019, 2021, 2021) have the advantage of internal consistency, being made using a single instrument and, in many cases, providing time resolution sufficient to monitor the heliocentric variation of the water production.   Where necessary, we used other published production rate data as listed in Table \ref{properties}.   

Production rates  measured as a function of heliocentric distance were interpolated to estimate the production rate at $r_H$ = 1 au, denoted $Q_{H_2O}(1)$, and listed in Table \ref{properties}. Some comets display significant asymmetry, such that $Q_{H_2O}(1)$ differs before and after perihelion.  In these cases we list the average value.  We estimate that the values of $Q_{H_2O}(1)$ in the table are accurate to no better than a factor of $\sim$2.  The  mass production rate at 1 au, $\dot{M}(1)$ kg s$^{-1}$, is related to $Q_{H_2O}(1)$ by $\dot{M}(1) = \mu m_H Q_{H_2O}(1)$, where $\mu$ = 18 is the molecular weight of the water molecule, and $m_H = 1.67\times10^{-27}$ kg is the mass of the hydrogen atom.  For reference, we note that a production rate $Q_{H_2O}(1) = 10^{29}$ s$^{-1}$ corresponds to $\dot{M}$ = 3000 kg s$^{-1}$.  We assume that gas produced on the nucleus surface flows away at the thermal speed $V_{th} = (8 k T/(\pi \mu m_H))^{1/2}$, where $k = 1.38\times10^{-23}$ J K$^{-1}$ is the Boltzmann constant and $T$ is the temperature of the sublimating ice surface.  At 1 au, the temperature is depressed by sublimation to about $T$ = 200 K,  giving $V_{th}$ = 500 m s$^{-1}$.  The temperature and $V_{th}$ change only slightly with heliocentric distance as a result of buffering by sublimation.  We ignore solid matter (``dust'') expelled simultaneously from the comets because, although the dust and gas mass production rates may be comparable, the dust mass is dominated by large particles which are poorly coupled to the gas flow, travel at speeds $\ll V_{th}$, and so carry only a small fraction of the outflow momentum. 

\section{RESULTS}

\subsection{Nucleus Radii}
\label{nucleus}

We use the production rate and non-gravitational acceleration data to estimate the radii of the nuclei in two ways. 

\textbf{a) Radius from Total Production Rate: }

In equilbrium with sunlight, sublimation of cometary ice drives a mass flux, $f_s$ (kg m$^{-2}$ s$^{-1}$), from the surface.  Measurements of  comets show that sublimation from the nucleus night-side is weak because surface temperatures there are very low.  Accordingly, we represent the nucleus as a sphere, sublimating only from the day-side hemisphere, and we use the energy balance equation to calculate $f_s$;

\begin{equation}
\frac{L_{\odot} (1-A_B)}{4\pi r_H^2} = 2\left[\epsilon \sigma T^4 + f_s(T) H(T)\right].
\label{energy}
\end{equation}

\noindent Here, the term on the left is the absorbed solar power ($L_{\odot} = 4\times10^{26}$ W is the solar luminosity, $A_B$ = 0.04 the assumed Bond albedo), the first term on the right accounts for radiation cooling ($\epsilon$ = 0.9 is the assumed emissivity, $\sigma = 5.67\times10^{-8}$ W m$^{-2}$ K$^{-4}$ is the Stephan-Boltzmann constant and $T$ is the effective temperature) and the second term represents energy consumed in sublimating ice the latent heat of which is $H = 2\times10^6$ J kg$^{-1}$).   Equation \ref{energy} is solved using the Clausius-Clapeyron equation for the pressure vs.~temperature along the sublimation phase change boundary.  At $r_H$ = 1 au, for  sublimation averaged over the dayside of a spherical ice nucleus, we find $f_s = 2.1\times10^{-4}$ kg m$^{-2}$ s$^{-1}$. 

The  mass loss rate, $\dot{M}$ (kg s$^{-1}$),  is then

\begin{equation}
\dot{M} = 2 \pi  r_n^2 f_A f_s
\label{emdot}
\end{equation}

\noindent where $r_n$ is the nucleus radius.  The factor $f_A$ represents the fraction of the surface area of the nucleus that contributes to the sublimation flux.  Empirically, $f_A$ is a decreasing function of nucleus radius and approaches unity on sub-kilometer short-period comets (Jewitt 2021a).  (This trend is probably a result of observational bias favoring the detection of small cometary nuclei having large active fractions over those with less active surfaces).  Values $f_A >$ 1 are possible if ice sublimates both from the nucleus and from icy grains ejected from the nucleus.      From Equation \ref{emdot} we obtain the nucleus radius

\begin{equation}
r_1 = \left(\frac{\dot{M}}{2 \pi f_Af_s}\right)^{1/2}
\label{rn1}
\end{equation}

\noindent and the nucleus mass

\begin{equation}
M_n(1) = \rho_n \left(\frac{2}{9\pi}\right)^{1/2} \left(\frac{\dot{M}}{f_A f_s}\right)^{3/2}.
\end{equation}

\noindent As a starting point, and in the absence of evidence regarding $f_A$ on the long-period nuclei, we take $f_A$ = 1. Values of $r_1$ are listed for each comet in Table \ref{derived}.    If $f_A <$ 1, as in most short-period comets \citep{Ahearn95}, then $r_1$ gives an under-estimate of the true radius.

\textbf{b) Radius from Non-Gravitational Acceleration: } Non-gravitational acceleration, $\alpha_{NG}$, is the result of anisotropic mass loss from sublimating ices on the nucleus. We use it to obtain a second estimate of the mass of the nucleus.  

The force on the nucleus is $k_R \dot{M} V_{th}$, where $V_{th}$ is the outflow speed and $0 \le k_R \le 1$ is a dimensionless constant expressing the fraction of the outflow momentum that goes into accelerating the nucleus (for isotropic outgassing, $k_R$ = 0, while for perfectly collimated outgassing $k_R$ = 1).     Then, force balance on a spherical nucleus of  density $\rho_n$ gives a second relation for the nucleus radius

\begin{equation}
r_2 = \left(\frac{3 k_R \dot{M} V_{th}}{4 \pi \rho_n \alpha_{NG}}\right)^{1/3}
\label{rn2}
\end{equation}

\noindent and the nucleus mass is

\begin{equation}
M_n(2) = \frac{k_R \dot{M} V_{th}}{\alpha_{ng}}.
\end{equation}

\noindent  We take the measured values of $\dot{M}$ and $\alpha_{NG}$ from Table \ref{properties} and, as noted above, we adopt $\rho_n$ = 500 kg m$^{-3}$ and $V_{th}$ = 500 m s$^{-1}$.  Measurements from 67P/Churyumov-Gerasimenko give $k_R$ = 0.5 (\cite{Jewitt20}), which we assume to apply to the long-period comets. The resulting values of $r_2$ are also listed for each comet in Table \ref{properties}.

\begin{figure}
\epsscale{0.75}
%\plotone{SB.pdf}
%\plotone{images.pdf}
%\plotone{dec28.pdf}
%\plotone{radius_radius.pdf}
\plotone{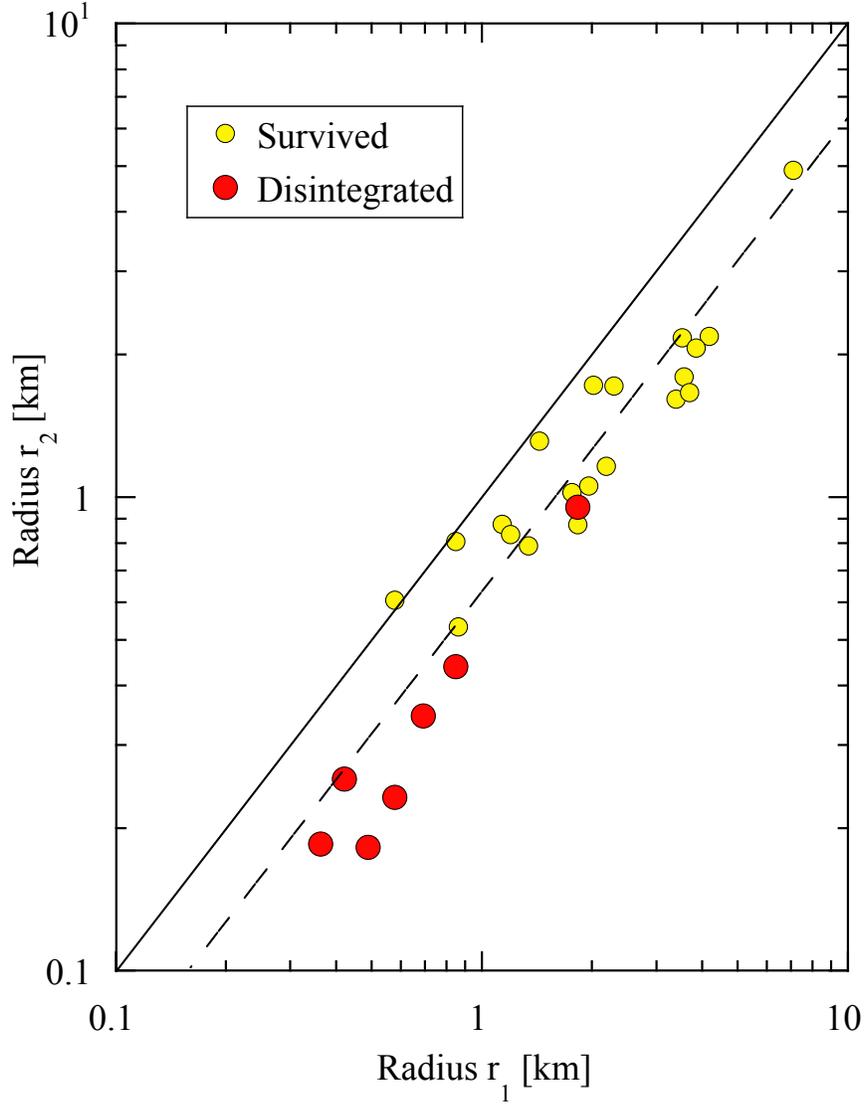}
\caption{Nucleus radii independently estimated from the production rate ($r_1$, from Equation \ref{rn1}) and the non-gravitational acceleration ($r_2$, from Equation \ref{rn2}).  Yellow and red-filled circles denote nuclei that survived and which were destroyed by the perihelion passage, respectively.  The solid line shows $r_1 = r_2$. The dashed line indicates $r_1 = (5/3) r_2$. \label{radius_radius}}
\end{figure}

The two estimates of the nucleus radii are compared in Figure \ref{radius_radius}.  Ideally, we would find $r_1 = r_2$ (indicated in the Figure by the solid diagonal line) but the comets are better described by $r_1 \sim (5/3) r_2$ (shown as a dashed, black line in the figure).    \cite{Sosa11} found a similar result  and interpreted it to mean that the long-period comets are hyperactive ($f_A >$ 1 in Equation \ref{rn1} acts to reduce $r_1$), allowing substantial sublimation from icy grains in the coma in addition to ice in the nucleus.  However, the assumption of hyperactivity is only one of several possible reasons for the difference between $r_1$ and $r_2$.  Parameters  $k_R$, and $\rho_n$  in Equation \ref{rn2} might also be different from the values assumed and radius estimate $r_2$ can be increased by increasing $k_R$ and/or decreasing $\rho_n$.  To consider one example,  $r_1$ and $r_2$ could be brought into  agreement if the bulk nucleus density in Equation \ref{rn2} were arbitrarily reduced by a  factor $(5/3)^3 \sim$ 5 to $\rho_n \sim$ 100 kg m$^{-3}$ instead of 500 kg m$^{-3}$, as assumed. This lower density is by no means ruled out by physics and would still be consistent with values measured in several short-period comets (e.g.~6P/d'Arrest, 19P/Borrelly)  \citep{Groussin19}. However, rather than make alternative, weakly justified guesses for some of the parameters in Equations \ref{rn1} and \ref{rn2}, we conservatively   choose the average radius, $\overline{r_n} = (r_1 + r_2)/2$, as our best estimate of the nucleus radius, and we take the difference between  radii $r_1$ and $r_2$ as a crude measure of the intrinsic radius uncertainty.   We feel that this is a good procedure because, if we instead followed \cite{Sosa11} by assuming that $f_A >$ 1, the  conclusions  to be reached (described in Section \ref{discussion}) would be changed only by becoming stronger.  The mean radii are listed in column 4 of Table \ref{derived}.

Figure \ref{radius_radius} shows that the disintegrating long-period comets (filled red circles representing C/2001 A2, C/2010 X1, C/2012 S1, C/2017 E4, C/2019 Y4, C/2020 F8 and C/2021 A1) have smaller nuclei, on average, than those that survive perihelion (filled yellow circles).  Six of the seven disintegrating comets are sub-kilometer bodies  while the seventh (C/2001 A2 (LINEAR)) is only slightly larger at $\overline{r_n} = 1.4\pm0.4$ km (Table \ref{properties}).  The median radius of the disintegrating long-period nuclei is 0.41 km (mean value 0.55$\pm$0.15 km, 7 objects) whereas that of surviving nuclei is 1.60 km (mean value 1.96$\pm$0.28 km, 20 objects).  The non-parametric KS test was used to assess the likelihood that the two radius distributions are drawn from the same population.  This test gave a statistic $D$ = 0.807, with an associated probability $p$ = 0.002, consistent with the visual impression that the surviving and disrupted radius distributions are distinct.   We also compared the distributions of perihelion distances of the surviving and disrupted comets.  The median perihelion distance of the disrupted comets is 0.48 au (mean value 0.48$\pm$0.09 au, 7 objects) while that of the survivors is 0.99 au (mean value 0.97$\pm$0.10 au, 20 objects).  The associated KS statistics are $D$ = 0.697 and $p$ = 0.007, indicating that the hypothesis that the two samples are drawn from the same parent population can again be rejected, although with less confidence.

\begin{figure}
\epsscale{0.75}
%\plotone{SB.pdf}
%\plotone{images.pdf}
%\plotone{dec28.pdf}
%\plotone{NGA_vs_DMBDT.pdf}
\plotone{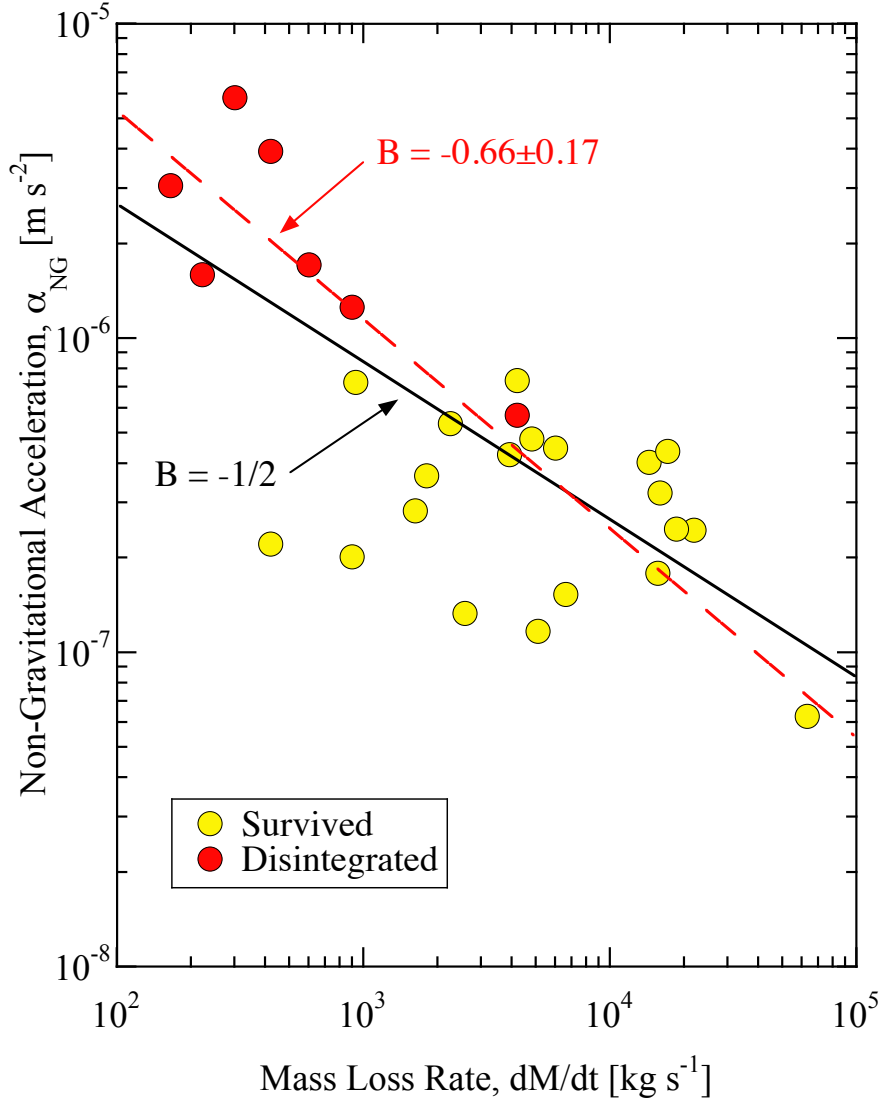}
\caption{Non-gravitational acceleration, $\alpha_{NG}$, as a function of the mass loss rate, $\dot{M}$, both referred to $r_H$ = 1 au, computed from the data in Table \ref{properties}.  Yellow and red-filled circles denote nuclei that survived and which were destroyed by the perihelion passage, respectively.  The solid black line has power-law slope $B$ = 0.5. The red dashed line shows the least-squares fit and its index. \label{nga_dmbdt}}
\end{figure}

Figure \ref{nga_dmbdt} compares the two measured quantities, the non-gravitational acceleration and the mass loss rate, both at 1 au.  The figure shows that $\alpha_{NG}(1)$ and $\dot{M}(1)$ are inversely related, and with a trend that is readily understood.  All else being equal, we expect that the mass loss rate should vary as $\dot{M}(1) \propto \overline{r_n}^2$ (Equation \ref{rn1}), while the non-gravitational acceleration should vary as $\alpha_{NG}(1) \propto \dot{M(1)}/\overline{r_n}^3 \propto 1/\overline{r_n}$ (Equation \ref{rn2}), giving $\alpha_{NG}(1) \propto \dot{M}(1)^{-1/2}$.  The black line in Figure \ref{nga_dmbdt} has slope -1/2 and evidently matches the data well.   A least-squares fit of a power law (dashed red line in the figure) to the data gives $\alpha_{NG}(1) \propto \dot{M}(1)^{B}$ with $B = -0.66\pm0.17$, consistent with this expectation within one standard deviation.  The color coding in Figure \ref{nga_dmbdt} also shows that comets with the largest non-gravitational accelerations and the weakest outgassing rates are the most likely to disintegrate, with an approximate separation between disintegrating and surviving comets at $\alpha_{NG}(1) \sim 10^{-6}$ m s$^{-2}$ and $\dot{M}(1) \sim 10^3$ kg s$^{-1}$.  By Equations \ref{rn1} and \ref{rn2}, these values correspond to nucleus radii $r_n$ = 0.5 to 0.9 km, consistent with the color-coding in Figure \ref{radius_radius} showing that subkilometer long-period comet nuclei disintegrate.

Published examples of well-characterized sub-kilometer long-period comets are few and far between; the compilation by \cite{Lamy04} lists only two.  C/1999 S4 (LINEAR) had $r_n$ = 0.45 km \citep{Altenhoff02}, $q$ = 0.765 au and disintegrated spectacularly at perihelion \citep{Weaver01}, consistent with our findings here. C/1983 J1 (Sugano-Saigusa-Fujikawa) had $r_n <$ 0.37 km \citep{Hanner87}, $q$ = 0.47 au but was observed only for a few weeks when near Earth so that its fate is unknown.      The paucity of well-studied small  nuclei relative to power-law extrapolations from larger sizes \citep{Bauer17} may itself be evidence for the efficient destruction of sub-kilometer long-period comet nuclei. 

\section{DISCUSSION}
\label{discussion}
Seven of the 27 LPCs in our sample  either fragmented or disintegrated, fates that are indicated in column 9 of Table \ref{derived} by the letters ``F'' and ``D'', respectively.  These descriptors are purely morphological; in fragmentation the comet splits into two or more discrete objects typically each retaining a cometary appearance, while in disintegration the comet assumes the appearance of an expanding, diffuse cloud, lacking an obvious source or other embedded structure.  The physical relationship between fragmentation and disintegration is unclear.  We assume that the former is a mild case of the latter and, for simplicity in the following discussion, we use the term ``disintegration'' to apply to both.

\subsection{Disintegration Mechanisms}
Given the results in Figures \ref{radius_radius} and \ref{nga_dmbdt}, what mechanisms could be  responsible for the disintegration of long-period nuclei?  

\textbf{Tidal Breakup:} The Roche radius of the Sun for a comet nucleus represented as a fluid body of density $\rho_n$ = 500 kg m$^{-3}$ is $\sim$10$^{-2}$ au. In our sample, only C/2012 S1 (ISON) approached the Sun closely enough ($q$ = 0.012 au) for tidal disruption to be possible.  Neither have the remaining comets passed within the Roche spheres of any planet, eliminating tidal breakup as a generally relevant mechanism for this study.

\textbf{Sublimation:} Sublimation erosion is typically $\sim$10 m per orbit (computed from Equation \ref{energy}), and thus is too slow to destroy a $\sim$1 km diameter nucleus in the $\sim$1 year spent in strong sublimation while close to the Sun.   Moreover, sublimation would naturally produce more steady erosion of the comet, not the catastrophic disintegrations as observed.  

\textbf{Collisional Disruption:} Interplanetary collisions are very rare.  Even in the relatively dense asteroid belt, the collisional disruption timescales of sub-kilometer bodies are measured in 100s  of Myr, such that the probability of a destructive collision in the $\sim$1 year spent by each comet near the Sun is negligible.  Moreover, the long-period comets have large orbits with random inclinations and disintegrate far above the ecliptic plane, where collisional disruption is even less likely.

\textbf{Confined Pressure Explosion:} \cite{Samarasinha01} invoked the build-up of gas pressure  in order to explain cometary disruption.  The problem with this mechanism is that the effect of heating by the Sun is confined to a thermal skin that is very thin compared to the radius of the nucleus.  For example, cometary material has very small diffusivity, resulting in a  thermal skin depth ($\sim$10$^{-2}$ m) that is very small compared to the nucleus radius. Unless the full body of the nucleus is permeated by large interconnected voids (but still sealed from the vacuum of surrounding space), any gas pressure build-up must be confined to a thin surface shell incapable of disrupting of the whole nucleus.  Likewise, crystallization runaways, although potentially capable of triggering cometary outbursts \citep{Prialnik92}, are necessarily confined by the radial temperature gradient to a  surface shell with thickness $\ll r_n$.

\textbf{Rotational Instability:} Outgassing exerts a torque on the cometary nucleus capable of substantially changing the  spin on a timescale given by 

\begin{equation}
\tau_s = \left(\frac{16\pi^2}{15}\right)\left(\frac{\rho_n r_n^4}{k_T V_{th} P} \right)\left(\frac{1}{\overline{\dot{M}}} \right).
\label{tau_s}
\end{equation}

\noindent Here, $P$ is the instantaneous spin-period, $k_T$ is the dimensionless moment arm, equal to the fraction of the outflow momentum that exerts a torque on the nucleus and $\overline{\dot{M}}$ is the average value of the mass loss rate following the comet in its orbit \citep{Jewitt21a}.  For a weakly cohesive nucleus, the end-state of spin-up is rotational disruption into fragments which are themselves subject to rapid disintegration because of the strong dependence of $\tau_s$ on $r_n$ (Equation \ref{tau_s}) and because of the sudden exposure of previously buried volatiles.  

To calculate $\overline{\dot{M}}$, we use the facts that water is the dominant volatile and that strong sublimation of water ice is  restricted to heliocentric distances $r_H \lesssim$ 3 au. Species more volatile than water (notably CO, CO$_2$) can sublimate at lower temperatures and larger distances but their abundances are poorly constrained in our sample of comets and, in the interests of simplicity, we do not consider them here. Their inclusion would only strengthen our conclusions by amplifying the importance of outgassing torques relative to our water ice calculation.   Accordingly, we computed the average mass loss rate for each comet using

\begin{equation}
\overline{\dot{M}} = \frac{\int_{t_0}^{t_0+\Delta t} \dot{M}(r_H(t)) dt}{\Delta t}
\end{equation}

\noindent where $t_0$ is the pre-perihelion time at which $r_H$ = 3 au and $\Delta t$ is the time spent with $r_H \le$ 3 au.  The instantaneous mass loss rate $\dot{M}(r_H(t))$ is computed by solving Kepler's law for $r_H(t)$ and using Equation \ref{energy} to find $\dot{M}(r_H(t)) = 2 \pi f_s(r_H(t)) \overline{r_n}^2$. The average, listed in Table \ref{derived}, is then substituted into Equation \ref{tau_s} in order to calculate the characteristic spin-up time, $\tau_s$.  

We again used  values of the quantities $k_T$, $\rho_n$ and $P$ taken from the cometary literature.  The median dimensionless moment arm, $k_T = 0.007$, is determined from measurements of the nuclei of short-period comets \citep{Jewitt21a}.  Likewise, the average nucleus density, $\rho_n = 480\pm220$ kg m$^{-3}$, is known only for the nuclei of short-period comets \citep{Groussin19}.  We adopt $P$ = 15 hours, equal to the median rotation period measured in short-period comets \citep{Jewitt21a}.   While there are no clear reasons to think that the median $k_T$, $\rho_n$ and $P$ should be different in the long-period vs.~short-period comets, we are aware of this possibility and eagerly await direct measurements of these parameters in the former population.  Given the uncertainties in these quantities, it is obvious that Equation \ref{tau_s} can provide, at best, a value of $\tau_s$ accurate to no better than order of magnitude.

\begin{figure}
\epsscale{0.75}
%\plotone{SB.pdf}
%\plotone{images.pdf}
%\plotone{dec28.pdf}
%\plotone{disint_plot.pdf}
\plotone{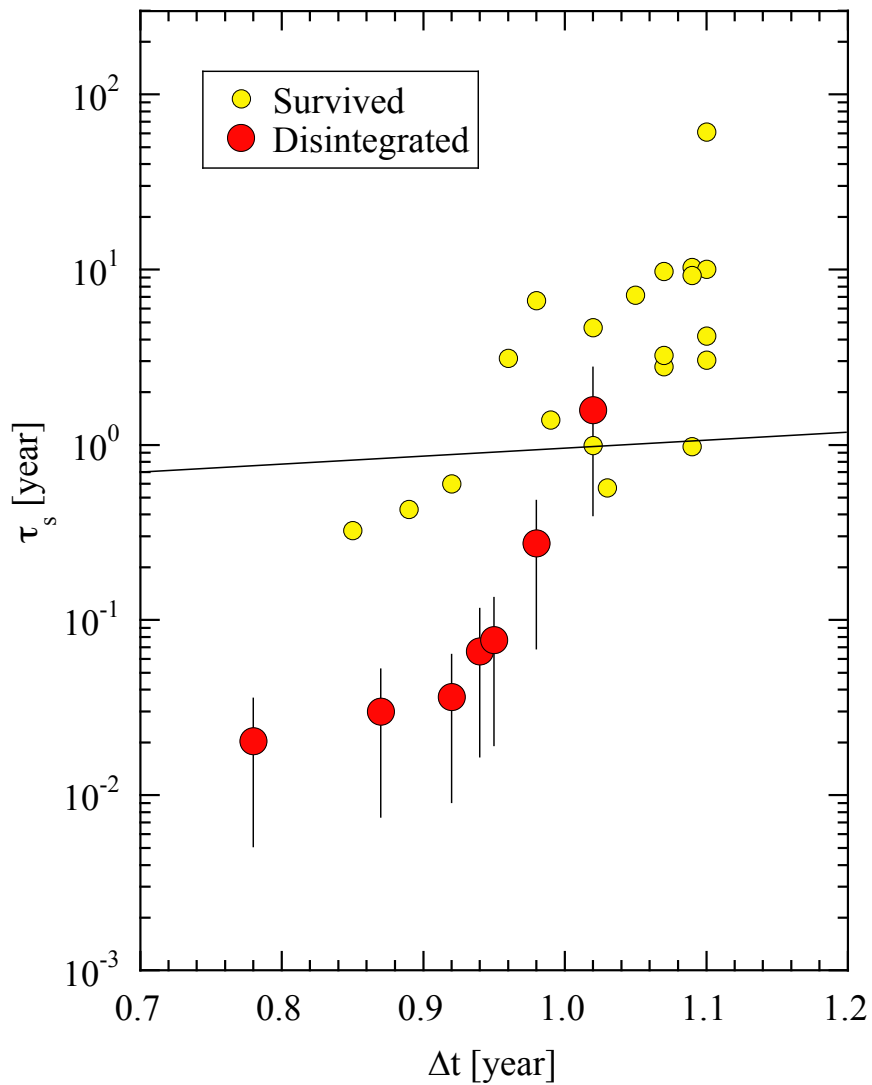}
\caption{Sublimation timescale, $\tau_s$ (Equation \ref{tau_s}) as a function of the time spent with $r_H <$ 3 au.  Yellow and red-filled circles denote nuclei that survived and which were destroyed by the perihelion passage, respectively.  The solid black line marks $ \tau_s = \Delta t$. \label{disint}}
\end{figure}

We set a simple criterion for judging the importance of spin-up torques by comparing $\tau_s$ with $\Delta t$, defined above as the time spent by each comet with $r_H \le$ 3 au.   If $\tau_s < \Delta t$ then sublimation torques can substantially modify the nucleus spin within a single perihelion passage of the comet, potentially leading to rotational instability and breakup.  Otherwise, the outgassing torques are too weak to trigger rotational breakup, at least within a single perihelion passage.  Values of $\Delta t$ and $\tau_s$ are listed in  Table \ref{derived} and plotted for convenience in Figure \ref{disint}.  The table and figure show that six of the seven disintegrated comets in our sample  had $\tau_s < \Delta t$, consistent with rotational breakup as the cause of their destruction.  The seventh (C/2001 A2) also satisfies this inequality within the error bar on $\tau_s$.   Those comets which did not breakup or disintegrate have $\tau_s > \Delta t$, again with some ambiguous cases close to the $\tau_s = \Delta t$ line.  Again, while emphasizing the (necessarily) order of magnitude nature of the treatment offered in Section \ref{nucleus} the basic result, that the disintegrating nuclei are those with the shortest spin-up times, is remarkable.

\begin{figure}
\epsscale{0.75}
%\plotone{SB.pdf}
%\plotone{images.pdf}
%\plotone{dec28.pdf}
%\plotone{emdot_vs_rn.pdf}
\plotone{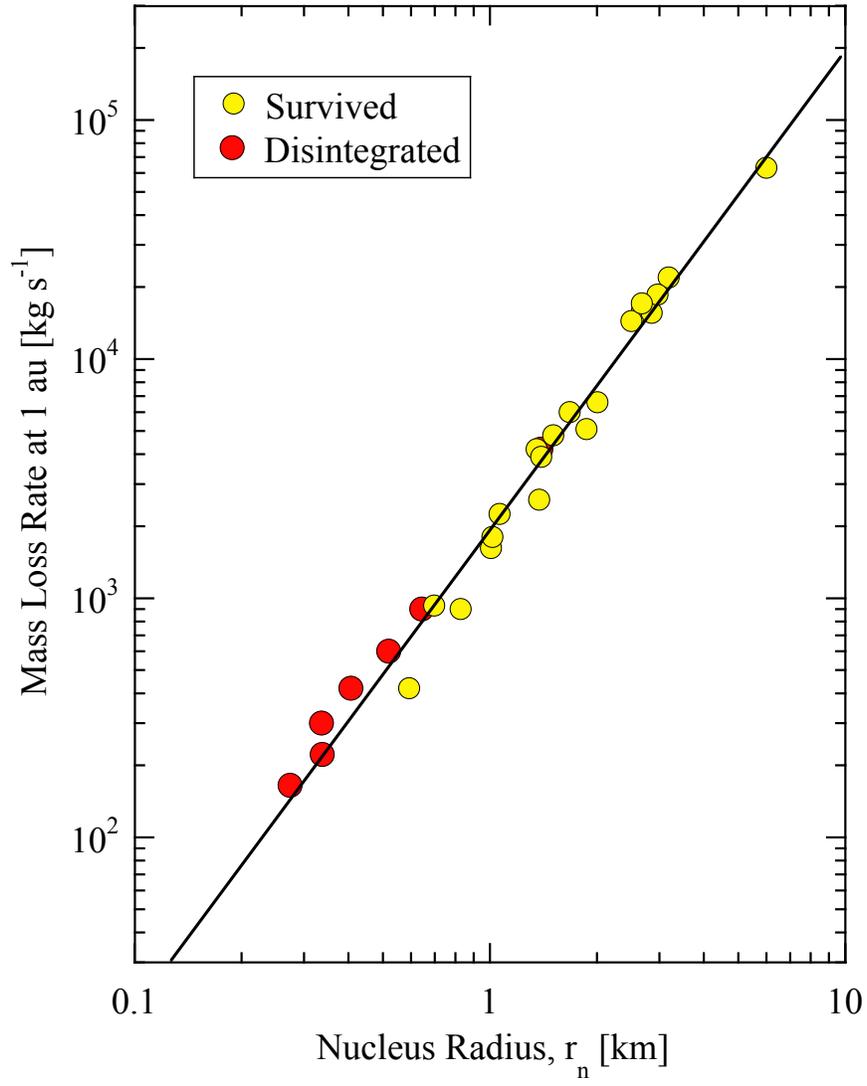}
\caption{Mass loss rate vs.~nucleus radius.  The solid black line shows $\dot{M}(1) = 1850 \overline{r_n}^2$. \label{emdot_vs_rn}}
\end{figure}

Figure \ref{emdot_vs_rn} shows the mass loss rate at $r_H$ = 1 au vs.~the nucleus radius, $\overline{r_n}$.   The solid black line in the figure shows $\dot{M}(1) = 1850 \overline{r_n^2}$, with $\overline{r_n}$ in km and $\dot{M}(1)$ in kg s$^{-1}$.   Substituting for $\dot{M}(1)$ in Equation \ref{tau_s} and setting  $\tau_s = \Delta t$ = 1 year, we calculate the critical radius below which the average long-period comet is susceptible to rotational break-up in a single perihelion passage as $r_n \sim$ 1 km, in agreement with the observation that the disintegrating long-period comets are sub-kilometer objects.  

\subsection{Relation to the Oort Fading Parameter}

Long-period comets with reciprocal semimajor axes $a^{-1} < 10^{-4}$ au$^{-1}$ are known as ``Oort spike'' comets, some but not all of which are dynamically new objects making their first pass through the planetary region.  \cite{Oort50} found that a purely dynamical model of comet delivery from a distant reservoir predicts a larger flux of returning objects, relative to dynamically new comets,  than is observed. He introduced an ad hoc ``fading parameter'' to bring the dynamical model into agreement with the data.  The need for this fading parameter has since been confirmed many times (e.g.~\cite{Wiegert99,Levison02,Neslusan06}) but the physical cause of the fading  remains unknown.  \cite{Oort50} and others conjecture that it is due to ``surface aging'' in response to insolation. \cite{Levison02} concluded from the small number of detections of returning objects in ground-based surveys that a majority of dynamically new comets are destroyed, not merely faded. 
   
%%%
The present work shows that disintegration, by removing comets from the observable sample, is a significant fading mechanism.  The destruction is size dependent, preferentially afflicting small nuclei, and also perihelion distance dependent, being more probable at small distances than at large.   In contrast, published models of the fading parameter instead assume a survival probability that is independent of nucleus size (\cite{Wiegert99,Levison02}).  \cite{Wiegert99} examined a sample of comets having median perihelion $q \sim$ 1 au, comparable to that of our sample in Table \ref{list}.  As one of several possible solutions, they found a fading law in which the comets are divided into two groups and supposed that some 95\% of long-period comets are destroyed within the first six perihelion passages while the remaining 5\% survive indefinitely.  We conjecture that this empirical division into two groups is an artifact of size-dependent rotational disruption; 95\% of the long-period comets are sub-kilometer objects subject to rotational disruption in a few orbits while 5\% are  larger and can resist disruption for a much longer time.  

The implication that only 5\% of long-period comet nuclei have $r_n >$ 1 km at first appears at odds with the radius distribution of the nuclei listed in Table \ref{derived}, where 17 of the 27 nuclei (62\%) are larger than 1 km.  However, our sample is highly observationally biased against the inclusion of small nuclei  because they produce too little H$_2$O to be detected in the flux-limited Lyman-$\alpha$  data from SWAN.    (The largest comets are also excluded because their non-gravitational accelerations are too small to be measured). As a result, the data in Table \ref{derived} severely underestimate the abundance of small LPC nuclei, and cannot be used to assess the intrinsic size distribution of the nuclei.  

Cometary fading has been reported to extend to at least   $r_H \sim$ 10 au \citep{Krolikowska19, Kaib22}, far beyond the region where outgassing torques from sublimating water ice can alter the nucleus spin.  Moreover, a growing number of observations show activity in distant comets (e.g.~$\sim$20 - 25 au in the case of C/2014 UN271 Bernardinelli-Bernstein \citep{Farnham21} and even 35 au in the case of C/2017 K2 (PANSTARRS) \citep{Jewitt21b}).   Fragmentation has also been inferred at very large distances, for example in comets C/2002 A1 and A2, reported to have split from a common parent when inbound at $r_H \sim$ 22.5 au \citep{Sekanina03}.   Water ice is involatile beyond 5 or 6 au and the sublimation of a more volatile material, perhaps carbon monoxide (CO) ice, is a leading candidate for driving this distant cometary activity.  Could fading at large distances be due to torques from CO sublimation?  

A definitive answer to this question cannot be reached given our limited knowledge of the surface properties of distant comets, or calculated from first principles.  However, order of magnitude scaling considerations strongly suggest an answer in the negative, as follows.  To first order, the ratio of the timescale for spin-up of a given body through sublimation of CO to that for spin-up through sublimation of water ice is (c.f.~Equation \ref{tau_s})

\begin{equation}
\frac{\tau_s(\textrm{CO})}{\tau_s(\textrm{H}_2\textrm{O})} \sim \frac{\dot{M}(\textrm{H}_2\textrm{O}) }{\dot{M}(\textrm{CO}) } \frac{ V_{th}(\textrm{H}_2\textrm{O})}{ V_{th}(\textrm{CO})}.
\label{ratio}
\end{equation}

\noindent We compare the sublimation of CO at 10 au to that of H$_2$O at 1 au.  Fortunately, measurements of the two ratios on the right hand side of Equation \ref{ratio} are available for the long-period comet C/1995 O1 (Hale-Bopp).  There, $\dot{M}(\textrm{H}_2\textrm{O}, r_H = 1 \textrm{ au}) / \dot{M}(\textrm{CO}, r_H = 10 \textrm{ au}) \sim 10^3$ (Figure 5 of \cite{Biver02}) and $V_{th}(\textrm{H}_2\textrm{O}, r_H = 1 \textrm{ au})/ V_{th}(\textrm{CO}, r_H = 10 \textrm{ au}) \sim$ 3 (figure 4a of \cite{Biver02}), giving $\tau_s(\textrm{CO}, r_H = 10 \textrm{ au}) / \tau_s(\textrm{H}_2\textrm{O}, r_H = 1 \textrm{ au}) \sim$ 3000.  This factor of 3000 is partly compensated by the longer time spent in the CO sublimation zone.  C/1995 O1 had $r_H <$ 30 au for  27 years  (compared with $\Delta t \sim$ 1 year for the comets in Table 1) and could have sublimated CO the entire time. Still, based on this scaling argument, spin-up timescales for a given object at 10 au remain two orders of magnitude larger than at 1 au, given C/Hale-Bopp-like outgassing behavior.  Substantial spin-up due to CO torques  seems unlikely unless  the CO/H$_2$O ratios in other LPCs are much larger than measured in C/1995 O1.  On this basis we conclude that, while important in the water sublimation zone, spin-up destruction is not an obvious cause of spin-up or fading in any but the tiniest comets beyond it.   Oort's fading parameter thus seems likely to have several physical origins of which rotational disruption is only one.

%%%The last two are problem paragraphs needing attention.

For LPCs with $q \lesssim$ 3 au, we offer three predictions that will be observationally testable in the foreseeable future given improved population data.  First, LPC nuclei larger than a few km in radius should rarely disrupt or disintegrate, unless by another process (e.g.~tidal disruption, as may be the case for C/2012 S1 (ISON)). Second, the size distribution of LPCs should be flattened at radii $r_n \lesssim$ 1 km, relative to its value at larger radii, owing to the selective loss of small nuclei through rotational instability.  Third, accurately bias-corected data should show that the size distributions of long-period comet nuclei  vary with $q$,  reaching ``primordial'' values only in the outer solar system where mass loss is negligible.  

Lastly, it is reasonable to expect that the imprints of rotational disruption might be found in the Damocloid population, to the extent that these objects (inactive bodies with Tisserand parameters $T_J \le$ 2) are remnants of formerly active long-period comets.  Measurements indeed show a flatter size distribution \citep{Kim14},  consistent with the preferential destruction of smaller Damocloids, but the available sample is small and undoubtedly subject to its own biases.  Future work on these objects may also be revealing.

\clearpage

\section{SUMMARY}
We examine a sample of 27 long-period comets for which both non-gravitational accelerations and water production rates are available.  Using these two measured quantities we are able to estimate the nucleus sizes, and so to explore the systematics of this population.  Seven of the 27 comets ($\sim$25\%) fragmented or disintegrated.

\begin{itemize}
\item The disintegrating cometary nuclei have systematically smaller radii (median 0.4 km, 7 objects)  than those that survive in proximity to the Sun (1.6 km, 20 objects).    %The median perihelion distances of the disintegrating and surviving comets are $q$ = 0.5 au and 1.0 au, respectively.

\item The disintegrating comets have smaller  perihelion distance (median 0.5 au,  7 objects) than those surviving (1.0 au, 20 objects).  

\item These size and perihelion distance trends are both  consistent with nucleus disintegration through rotational instability, triggered by outgassing torques from sublimating water ice.   Specifically, the timescale for outgassing torques to change the spin of sub-kilometer nuclei is less than the time spent in strong sublimation.  

\item Rotational disruption is a cause of the ``fading'' required to fit the orbital semimajor axis distribution of long-period comets.

\end{itemize}

\clearpage

\acknowledgments
I thank the anonymous referee for highlighting the importance of fading at large distances, Yoonyoung Kim for additional comments on the manuscript and Man-To Hui for advice about the vagaries of JPL Horizons.  Based in part on observations made under GO 16929 with the NASA/ESA Hubble Space Telescope, obtained at the Space Telescope Science Institute, operated by the Association of Universities for Research in Astronomy, Inc., under NASA contract NAS 5-26555.


%% in parentheses, after the keyword, but they will not be verified.

%{\it Facilities:}  \facility{}.

\clearpage


\begin{deluxetable}{lrrrrrrrrr}
\tabletypesize{\scriptsize}
%\rotate
\tablecaption{Comet Sample 
\label{list}}
\tablewidth{0pt}
\tablehead{\colhead{Comet} & \colhead{$a$\tablenotemark{a}}  & \colhead{$e$\tablenotemark{b}} & \colhead{$i$\tablenotemark{c}}  & \colhead{$q$\tablenotemark{d}} & \colhead{$T_P$\tablenotemark{e}}  & \colhead{S/NS\tablenotemark{f}}     }

\startdata
C/2000	WM1	(LINEAR)		&	1877		& 0.9997077	&	72.6		&0.549	&	2452297.3 & NS\\
C/2001	A2-A	(LINEAR)		&	971		& 0.9991978	&	36.5		&0.779	&	2452054.0 & NS  \\
C/2001	Q4	(NEAT)		&	16725		& 0.9999426	&	99.6		&0.960	&	2453141.5 & S  \\
C/2002	T7	(LINEAR)		&	47471		& 0.9999870	&	160.6 		&0.615	&	2453118.5 & S  \\
C/2002	X5	(Kudo-Fujikawa)	&	1119		& 0.9998281	&	94.2		& 0.192	&	2452668.5 & NS  \\
C/2003	K4	(LINEAR)		&	29199		& 0.9999651	&	134.2		& 1.021	&	2453292.3 & S  \\
C/2004	Q2	(Machholz)	&	2528			& 0.9995221	&	38.6		& 1.208	&	2453395.5 & NS  \\
C/2009	P1	(Garradd)		&	2384		& 0.9993522	&	106.2		&1.544	&	2455919.3 & NS  \\
C/2010	X1	(Elenin)		&	48388		& 0.9999900	&	1.8		&0.482	&	2455815.2 & S  \\
C/2012	K1 (PANSTARRS)	&	26070			& 0.9999597	&	142.4		&1.051	&	2456897.3 & S  \\
C/2012	S1	(ISON)		&	-144820	& 1.0000001	&	62.2		&0.012	&	2456625.3 & S  \\

C/2012	X1	(LINEAR)		&	145		& 0.9889527	&	44.4		&1.597	&	2456710.3 & NS \\
C/2013	US10	(Catalina)	&	19030			& 0.9999569	&	148.9		&0.820	&	2457342.3 & S \\
C/2013	X1 (PANSTARRS)	&	3804 			& 0.9996529	&	163.2		&1.320	&	2457499.3 & NS \\
C/2014	E2	(Jacques)		&	807		& 0.9991826	&	156.4		&0.664	&	2456841.0 & NS \\

C/2014	Q1 (PANSTARRS)	&	841			& 0.9996251	&	43.1		&0.315	&	2457210.0 & NS \\
C/2014	Q2	(Lovejoy)		&	502		& 0.9974320	&	80.3		&1.290	&	2457052.5 & NS \\
C/2015	ER61	(PANSTARRS)	&	718		& 0.9985315	&	6.3		&1.054	&	2457883.5 & NS \\
C/2015	G2	(MASTER)	&	6103 			& 0.9998724	&	147.6		&0.779	&	2457166.2 & NS \\

C/2015	V2	(Johnson)		&	2525 		& 0.9999546	&	49.9		&1.631	&	2457916.8 & NS \\
C/2017	E4	(Lovejoy)		&	821		& 0.9994044	&	88.2		&0.489	&	2457866.8 & NS \\
C/2017	T2 (PANSTARRS)	&	36830	 		& 0.9999560	&	57.2		&1.619	&	2458974.5 & S \\
C/2019	Y1	(ATLAS)		&	209		& 0.9959967	&	73.3		&0.836	&	2458924.0 & NS \\
C/2019	Y4-B	(ATLAS)		&	501		& 0.9994983	&	45.4		&0.251	&	2459000.5 & NS \\
C/2020	F8	(SWAN)		&	-4886		& 1.0000874	&	110.8		&0.427	&	2458997.0 & S \\
C/2020	S3	(Erasmus)		&	191		& 0.9978918	&	19.9		&0.403	&	2459196.3 & NS \\
C/2021	A1	(Leonard)		&	2028 		& 0.9996965	&	132.7	&0.616	&	2459582.8 & NS \\

\enddata

%% Text for table notes should follow after the \enddata but before
%% the \end{deluxetable}. Make sure there is at least one \tablenotemark
%% in the table for each \tablenotetext.

\tablenotetext{a}{Barycentric semimajor axis, in au}
\tablenotetext{b}{Eccentricity}
\tablenotetext{c}{Inclination, in degrees }
\tablenotetext{d}{Perihelion distance, in au }
\tablenotetext{e}{Mean Julian Date of Perihelion }
\tablenotetext{f}{S = Spike, NS = Non-Spike }

\end{deluxetable}

\clearpage

%%%%%%%%%%%%%%%%%%%%%%%%%%%%%%%%%%%%%%%%%
%%%%%%%%%%%%%%%%%%%%%%%%%%%%%%%%%%%%%%%%%
%%%%%%%%%%%%%%%%%%%%%%%%%%%%%%%%%%%%%%%%%

%

\begin{deluxetable}{lccrrrrrrr}
\tabletypesize{\scriptsize}
%\rotate
\tablecaption{Measured Properties 
\label{properties}}
\tablewidth{0pt}
\tablehead{\colhead{Comet} & \colhead{$A_1$\tablenotemark{a}}  & \colhead{$A_2$\tablenotemark{a}} & \colhead{$A_3$\tablenotemark{a}}  & \colhead{$\alpha_{NG}$\tablenotemark{b}} & \colhead{$Q_{H_2O}(1)$\tablenotemark{c}}   & Source\tablenotemark{d}  }

\startdata
C/2000 WM1	(LINEAR)	&	5.8e-09	&	-8.0e-11	&	0	&	1.2e-07	&	1.7e+29			&	C19	\\
C/2001 A2	(LINEAR)	&	-2.1e-08	&	1.9e-08	&	0	&	5.7e-07	&	1.4e+29				&	C08	\\
C/2001 Q4	(NEAT)	&	1.6e-08	&	4.8e-10	&	4.3e-10	&	3.2e-07	&	5.3e+29		&	C09	\\
C/2002 T7	(LINEAR)	&	1.2e-08	&	9.6e-10	&	-1.8e-09	&	2.4e-07	&	7.3e+29			&	C19, S20	\\
C/2002 X5	(Kudo-Fujikawa)	&	2.6e-08	&	5.8e-09	&	0	&	5.4e-07	&	7.5e+28	&	C11	\\
C/2003 K4	(LINEAR)	&	8.1e-09	&	-3.6e-09	&	-5.6e-10	&	1.8e-07	&	5.2e+29		&	C19	\\
C/2004 Q2	(Machholz)	&	1.2e-08	&	-1.1e-09	&	-2.3e-09	&	2.5e-07	&	6.2e+29	&	C19	\\
C/2009 P1	(Garradd)	&	2.0e-08	&	-1.0e-09	&	0	&	4.0e-07	&	4.8e+29			&	C19	\\
C/2010 X1	(Elenin)	&	-4.8e-08	&	6.3e-08	&	0	&	1.6e-06	&	7.4e+27			&	S11	\\
C/2012 K1	(PANSTARRS)	&	2.2e-08	&	-1.6e-09	&	-2.6e-09	&	4.5e-07	&	2.0e+29	&	C19	\\
C/2012 S1	(ISON)	&	8.5e-08	&	5.8e-09	&	0	&	1.7e-06	&	2.0e+28			&	C19	\\
C/2012 X1	(LINEAR)	&	3.6e-08	&	2.2e-09	&	5.3e-09	&	7.3e-07	&	1.4e+29		&	L14	\\
C/2013 US10	(Catalina)	&	7.6e-09	&	6.2e-11	&	1.6e-10	&	1.5e-07	&	2.2e+29		&	C19	\\
C/2013 X1	(PANSTARRS)	&	2.0e-08	&	-4.1e-09	&	-7.3e-09	&	4.4e-07	&	5.7e+29	&	C19	\\
C/2014 E2	(Jacques)	&	2.1e-08	&	-2.8e-09	&	0	&	4.3e-07	&	1.3e+29			&	C19	\\
C/2014 Q1	(PANSTARRS)	&	7.9e-09	&	2.8e-09	&	-5.5e-09	&	2.0e-07	&	3.0e+28	&	C19	\\
C/2014 Q2	(Lovejoy)	&	1.3e-09	&	-1.5e-09	&	-2.4e-09	&	6.3e-08	&	2.1e+30		&	C19	\\
C/2015 ER61	(PANSTARRS)	&	5.6e-09	&	-3.5e-09	&	-4.6e-10	&	1.3e-07	&	8.6e+28	&	S20	\\
C/2015 G2	(MASTER)	&	1.2e-08	&	5.9e-09	&	-4.3e-09	&	2.8e-07	&	5.4e+28	&	C19	\\
C/2015 V2	(Johnson)	&	2.3e-08	&	-4.2e-09	&	-4.4e-09	&	4.8e-07	&	1.6e+29		&	C21	\\
C/2017 E4	(Lovejoy)	&	1.8e-07	&	-7.6e-08	&	0	&	3.9e-06	&	1.4e+28			&	F18 \\
C/2017 T2	(PANSTARRS)	&	3.6e-08	&	-7.1e-10	&	-3.2e-10	&	7.2e-07	&	3.1e+28		&	C21	\\
C/2019 Y1	(ATLAS)	&	1.1e-08	&	0	&	0	&	2.2e-07	&	1.4e+28				&	C21	\\
C/2019 Y4	(ATLAS)	&	2.9e-07	&	-9.1e-09	&	0	&	5.8e-06	&	1.0e+28			&	C21	\\
C/2020 F8	(SWAN)	&	1.5e-07	&	-2.6e-08	&	0	&	3.1e-06	&	5.5e+27				&	C21	\\
C/2020 S3	(Erasmus)	&	1.7e-08	&	6.3e-09	&	0	&	3.6e-07	&	6.0e+28			&	C21b	\\
C/2021 A1	(Leonard)	&	5.8e-08	&	-2.0e-08	&	1.1e-08	&	1.3e-06	&	3.0e+28			&	C22	\\
\enddata

%% Text for table notes should follow after the \enddata but before
%% the \end{deluxetable}. Make sure there is at least one \tablenotemark
%% in the table for each \tablenotetext.

\tablenotetext{a}{Orthogonal components of the non-gravitational acceleration (units au day$^{-2}$) from JPL Horizons}
\tablenotetext{b}{Total non-gravitational acceleration at $r_H$ = 1 au (units m s$^{-2}$) from Equation \ref{alpha}}
\tablenotetext{c}{Water production rate at $r_H$ = 1 au, molecules s$^{-1}$}
\tablenotetext{d}{C08 = \cite{Combi08}, C09 = \cite{Combi09}, C11 = \cite{Combi11}, C19 = \cite{Combi19}, C21 = \cite{Combi21}, C21b = \cite{Combi21b}, C22 = M. Combi (private communication), F18 = \cite{Faggi18}, S11 = D. Schleicher (private communication, cited in \cite{Li15}), S20 = \cite{Saki20} }
\end{deluxetable}

\clearpage

\clearpage


\begin{deluxetable}{lccrlrrrrr}
\tabletypesize{\scriptsize}
%\rotate
\tablecaption{Derived Properties  
\label{derived}}
\tablewidth{0pt}
\tablehead{\colhead{Comet} & $r_1$\tablenotemark{a} & $r_2$\tablenotemark{b} & $\overline{r}$\tablenotemark{c} & \colhead{$Q_{H_2O}$\tablenotemark{d}}  & $\overline{\dot{M}}$\tablenotemark{e} & \colhead{$\tau_s$\tablenotemark{f}} & \colhead{$\Delta t $\tablenotemark{g}}  & \colhead{Note\tablenotemark{h} }     }

\startdata

C/2000	WM1	(LINEAR)	&	2.0	&	1.7	&	1.9	&	1.2e+29	&	3628 	&	3.13	&	0.96	&		\\
C/2001	A2-A	(LINEAR)	&	1.8	&	1.0	&	1.4	&	7.3e+28	&	2188	 &	1.58	&	1.02	&	F	\\
C/2001	Q4	(NEAT)	&	3.6	&	1.8	&	2.7	&	2.2e+29	&	6691	   &	7.15	&	1.05	&		\\
C/2002	T7	(LINEAR)	&	4.2	&	2.2	&	3.2	&	4.7e+29	&	14263	&	6.65	&	0.98	&		\\
C/2002	X5	(Kudo-Fujikawa)	&	1.3	&	0.8	&	1.1	&	1.2e+29	&	3607	&	0.32	&	0.85	&		\\
C/2003	K4	(LINEAR)	&	3.5	&	2.2	&	2.8	&	2.1e+29	&	6252	&	9.79	&	1.07	&		\\
C/2004	Q2	(Machholz)	&	3.9	&	2.1	&	3.0	&	2.3e+29	&	6895	&	10.32	&	1.09	&		\\
C/2009	P1	(Garradd)	&	3.4	&	1.6	&	2.5	&	1.2e+29	&	3607	&	10.04	&	1.10	&		\\
C/2010	X1	(Elenin)	&	0.4	&	0.3	&	0.3	&	6.0e+27	&	180	&	0.07	&	0.94	&	D	\\
C/2012	K1	(PANSTARRS)	&	2.2	&	1.2	&	1.7	&	8.8e+28	&	2645	&	2.79	&	1.07	&		\\
C/2012	S1	(ISON)	&	0.7	&	0.3	&	0.5	&	1.1e+29	&	3306	&	0.02	&	0.78	&	D	\\
C/2012	X1	(LINEAR)	&	1.8	&	0.9	&	1.4	&	3.4e+28	&	1010	&	3.05	&	1.10	&		\\
C/2013	US10	(Catalina)	&	2.3	&	1.7	&	2.0	&	1.1e+29	&	3240	&	4.67	&	1.02	&		\\
C/2013	X1	(PANSTARRS)	&	3.7	&	1.7	&	2.7	&	1.7e+29	&	5140	&	9.31	&	1.09	&		\\
C/2014	E2	(Jacques)	&	1.8	&	1.0	&	1.4	&	8.3e+28	&	2501.0	&	1.38	&	0.99	&		\\
C/2014	Q1	(PANSTARRS)	&	0.8	&	0.8	&	0.8	&	3.4e+28	&	1010	&	0.43	&	0.89	&		\\
C/2014	Q2	(Lovejoy)	&	7.1	&	4.9	&	6.0	&	6.5e+29	&	19569	&	61.01	&	1.10	&		\\
C/2015	ER61 (PANSTARRS)	&	1.4	&	1.3	&	1.4	&	3.4e+28	&	1008	&	3.24	&	1.07	&		\\
C/2015	G2	(MASTER)	&	1.1	&	0.9	&	1.0	&	3.2e+28	&	973	&	0.99	&	1.02	&		\\
C/2015	V2	(Johnson)	&	2.0	&	1.1	&	1.5	&	3.8e+28	&	1154	&	4.18	&	1.10	&		\\
C/2017	E4	(Lovejoy)	&	0.6	&	0.2	&	0.4	&	1.1e+28	&	328	&	0.08	&	0.95	&	D	\\
C/2017	T2	(PANSTARRS)	&	0.9	&	0.5	&	0.7	&	7.4e+27	&	223	&	0.98	&	1.09	&		\\
C/2019	Y1	(ATLAS)	&	0.6	&	0.6	&	0.6	&	6.7e+27	&	202	&	0.57	&	1.03	&		\\
C/2019	Y4	(ATLAS)	&	0.5	&	0.2	&	0.3	&	1.3e+28	&	393	&	0.03	&	0.87	&	F	\\
C/2020	F8	(SWAN)	&	0.4	&	0.2	&	0.3	&	4.8e+27	&	143	&	0.04	&	0.92	&	D	\\
C/2020	S3	(Erasmus)	&	1.2	&	0.8	&	1.0	&	5.6e+28	&	1677	&	0.60	&	0.92	&		\\
C/2021	A1	(Leonard)	&	0.8	&	0.4	&	0.6	&	1.9e+28	&	577	&	0.27	&	0.98	&	D	\\

\enddata

%% Text for table notes should follow after the \enddata but before
%% the \end{deluxetable}. Make sure there is at least one \tablenotemark
%% in the table for each \tablenotetext.
\tablenotetext{a}{Nucleus radius from Equation \ref{rn1}, km}
\tablenotetext{b}{Nucleus radius from Equation \ref{rn2}, km}
\tablenotetext{c}{Mean nucleus radius, km}
\tablenotetext{d}{Average water production rate when $r_H \le$ 3 au} % DJ: check 2 vs 3 au
\tablenotetext{e}{Average mass production rate when $r_H \le $ 3 au, kg s$^{-1}$}
\tablenotetext{f}{Spin change timescale, from Equation \ref{tau_s}, in years}
\tablenotetext{g}{Elapsed time with $r_H \le $ 3 au, in years}
\tablenotetext{h}{D = Disintegrated, F = Fragmented}

%\tablenotetext{d}{Perihelion distance, in au }

\end{deluxetable}

\clearpage

%%%%%%%%%%%%%%%%%%%%%%%%%%%%%%%%%%%%%%%%%
%%%%%%%%%%%%%%%%%%%%%%%%%%%%%%%%%%%%%%%%%
%%%%%%%%%%%%%%%%%%%%%%%%%%%%%%%%%%%%%%%%%

%
\clearpage
%%%%%%%%%%%%%%%%%%%%%%%%%%%%%%%%%%%%%%%%%
%%%%%%%%%%%%%%%%%%%%%%%%%%%%%%%%%%%%%%%%%
%%%%%%%%%%%%%%%%%%%%%%%%%%%%%%%%%%%%%%%%%

\clearpage
%%%%%%%%%%%%%%%%%%%%%%%%%%%%%%%%%%%%%%%%%
%%%%%%%%%%%%%%%%%%%%%%%%%%%%%%%%%%%%%%%%%
%%%%%%%%%%%%%%%%%%%%%%%%%%%%%%%%%%%%%%%%%

\clearpage
%%%%%%%%%%%%%%%%%%%%%%%%%%%%%%%%%%%%%%%%%
%%%%%%%%%%%%%%%%%%%%%%%%%%%%%%%%%%%%%%%%%
%%%%%%%%%%%%%%%%%%%%%%%%%%%%%%%%%%%%%%%%%

\clearpage
%%%%%%%%%%%%%%%%%%%%%%%%%%%%%%%%%%%%%%%%%
%%%%%%%%%%%%%%%%%%%%%%%%%%%%%%%%%%%%%%%%%
%%%%%%%%%%%%%%%%%%%%%%%%%%%%%%%%%%%%%%%%%

%\clearpage
%%%%%%%%%%%%%%%%%%%%%%%%%%%%%%%%%%%%%%%%%%
%%%%%%%%%%%%%%%%%%%%%%%%%%%%%%%%%%%%%%%%%%
%%%%%%%%%%%%%%%%%%%%%%%%%%%%%%%%%%%%%%%%%%
%\begin{figure}
%\epsscale{0.9}
%%\plotone{SB.pdf}
%%\plotone{images.pdf}
%%\plotone{dec28.pdf}
%\plotone{H10_vs_rn.pdf}
%
%\caption{Absolute magnitude, $H_{10}$ (from JPL Horizons), as a function of nucleus radius, $r_n$ (from Table \ref{derived}).  The best fit (Equation \ref{H10fit}) is shown as a solid black line, with slope -3.8$\pm$0.6.  \label{H10_vs_rn}}
%\end{figure}
%

%
%

%
%
%

\end{document}